\documentclass[a4paper,3p,12pt]{elsarticle}
\usepackage{graphicx,dcolumn,bm,amsfonts,amssymb,mathtools,color,natbib}
\usepackage{tgtermes,mathrsfs}
\usepackage[colorlinks]{hyperref}

\biboptions{sort&compress} 

\newcommand{\ud}{\mathrm{d}}

\newcommand{\om}{\omega_{\textrm{\tiny BD}}}

\begin{document} 
\begin{frontmatter}

\title{On the structural stability of a simple cosmological model in 
\boldmath $R+\alpha R^{2}$ theory of gravity}

\author{Orest Hrycyna}
\ead{orest.hrycyna@ncbj.gov.pl} 
\address{Theoretical Physics Division, National Centre for Nuclear Research,  Ludwika Pasteura 7, 02-093 Warszawa, Poland}

\begin{abstract} 
The theory of gravity with a quadratic contribution of scalar curvature is investigated using a dynamical systems approach. The simplest Friedmann--Robertson--Walker metric is employed to formulate the dynamics in both the Jordan frame and the conformally transformed Einstein frame. We show that, in both frames, there are stable de Sitter states where the expansion of the Hubble function naturally includes terms corresponding to an effective dark matter component. Using the invariant center manifold, we demonstrate that, in the Einstein frame, there exists a zero-measure set of initial conditions that lead from an unstable to a stable de Sitter state. Additionally, the initial de Sitter state is associated with a parallelly propagated singularity. We show that the formulations of the theory in the Jordan frame and the Einstein frame are physically nonequivalent.
\end{abstract}

\end{frontmatter}

\section{Introduction and the model}

Sakharov was the first to notice that the Einstein-Hilbert action integral is the lowest order in an infinite series of corrections in curvature invariants \cite{Sakharov:1967pk}. Then it was noticed that the gravitational theory can be treated as an effective field theory with a cut-off scale \cite{Zeldovich:1967, Stelle:1977ry, Adler:1983}. The first $f(R)$ theory was investigated in \cite{Polievktov:1969}, and quadratic contributions to the Einstein field equations were considered in \cite{Ruzmaikina:1969}. Finally, the idea of inflation as a transient phenomenon in $R^{2}$ cosmology emerged \cite{Ginzburg:1971, Starobinsky:1980te, Weinberg:1983, Whitt:1984pd, Starobinsky:1987zz, Maeda:1985bq, Maeda:1987xf, Berkin:1990nu, Capozziello:1993xn, Ketov:2012jt}

Dynamical systems analysis has been widely used in cosmological applications since the seminal papers by Belinskii \cite{Belinskii:1985, Belinskii:1987}, and stability analysis of cosmological models is of utmost importance \cite{Hawking:1971bv, Barrow:1983rx, Szydlowski:1984, Coley:1992, Kokarev:2008ba}. In the present paper, we extend stability analysis by using the concept of structural stability \cite{Andronov:1937,Thom:book}.  

The most general action integral for the gravitational theory can be presented as an infinite series in curvature invariants \cite{Donoghue:1994dn, Donoghue:1995cz}

\begin{equation}
S = \int\ud^{4}x\sqrt{-g}\left(-2\phi_{0}\Lambda+\phi_{0}R+c_{1}R^{2}+c_{2}R_{\mu\nu}R^{\mu\nu}
+c_{3}R_{\mu\nu\alpha\beta}R^{\mu\nu\alpha\beta}+ \dots \right) + 16\pi\,S_{m}
\end{equation}
where $\phi_{0}$ is a dimensional constant, and $c_{1}$, $c_{2}$, $c_{3}$ are dimensionless constants of the theory. We work in units where the speed of light and the Planck constant are set to $c=\hbar=1$.

\subsection{Jordan frame}

We begin with the following truncated Sakharov action integral \cite{Sakharov:1967pk} for gravitational theory:
\begin{equation}
\label{eq:act_orig}
S_{g}=\int\ud^{4}x\sqrt{-g}\left(-2\phi_{0}\Lambda+\phi_{0}R+c_{1}R^{2}\right)
\end{equation}
where the constant $\phi_{0}$ and the cosmological constant $\Lambda$ have dimensions of $M^{2}$. $c_{1}$ is a dimensionless constant of the theory. 

We introduce the following redefinition of the dynamical variable:
\begin{equation}
\varphi=\phi_{0}+2c_{1}R\,
\end{equation}
which implies that the Ricci scalar can be expressed as
\begin{equation}
R=\frac{\varphi-\phi_{0}}{2c_{1}}\,.
\end{equation}
Substituting this into the action integral, we obtain
\begin{equation}
S_{g}=\int\ud^{4}x\sqrt{-g}\left(\varphi R -2\phi_{0}\Lambda-\frac{(\varphi-\phi_{0})^{2}}{4c_{1}}\right)\,.
\end{equation}
This is a special case of the Brans-Dicke theory with $\om\equiv0$, known as  the O'Hanlon theory \cite{ohanlon:1972}. The action integral can be presented as 
\begin{equation}
\label{eq:action_jf}
S_{g}=\int\ud^{4}x\sqrt{-g}\left(\varphi R -2 V(\varphi)\right)\,,
\end{equation}
where the scalar field potential function is
\begin{equation}
V(\varphi)=\phi_{0}\Lambda+\frac{(\varphi-\phi_{0})^{2}}{8c_{1}}=\phi_{0}\left(\Lambda+\tilde{\Lambda}\left(\frac{\varphi}{\phi_{0}}-1\right)^{2}\right)\,,
\end{equation}
and we have introduced the dimensionful constant
\begin{equation}
\tilde{\Lambda}=\frac{\phi_{0}}{8c_{1}}\,,
\end{equation}
which can be interpreted as the Einstein frame cosmological constant.

We must remember that the scalar field presented in the theory has a gravitational origin, and its ontology is completely different from that in particle physics.

Variation of the action integral \eqref{eq:action_jf} with respect to the metric tensor, $\delta S_{g}/\delta g^{\mu\nu}=0$, yields the field equations for the theory:
\begin{equation}
\varphi\left(R_{\mu\nu}-\frac{1}{2}g_{\mu\nu}R\right)+V(\varphi)g_{\mu\nu}+g_{\mu\nu}\Box\varphi-\nabla_{\mu}\nabla_{\nu}\varphi=0\,,
\end{equation}
where $\Box = g^{\alpha\beta}\nabla_{\alpha}\nabla_{\beta}$.
Variation of the action with respect to the scalar field, $\delta S_{g}/\delta\varphi=0$, gives
\begin{equation}
R=2 \frac{\partial V(\varphi)}{\partial\varphi}=2V'(\varphi)\,.
\end{equation}
Taking the trace of the field equations yields the scalar field equation of motion:
\begin{equation}
\Box\varphi +\frac{2}{3}\left(2V(\varphi)-\varphi V'(\varphi)\right)=0\,.
\end{equation}

We now consider the flat Friedmann--Robertson--Walker metric:
\begin{equation}
\ud s^{2} = -\ud t^{2} + a^{2}(t)\left(\ud x^{2}+\ud y^{2} + \ud z^{2}\right)\,,
\end{equation}
and obtain the following energy conservation condition:
\begin{equation}
\label{eq:en_JF}
H^{2} = -H \frac{\dot{\varphi}}{\varphi}+\frac{\phi_{0}}{\varphi}\left(\frac{\Lambda}{3}+\frac{\tilde{\Lambda}}{3}\left(\frac{\varphi}{\phi_{0}}-1\right)^{2}\right)\,,
\end{equation}
the acceleration equation:
\begin{equation}
\label{eq:accel_JF}
\dot{H}=-2H^{2}+2\frac{\tilde{\Lambda}}{3}\left(\frac{\varphi}{\phi_{0}}-1\right)\,,
\end{equation}
and the equation of motion for the scalar field:
\begin{equation}
\label{eq:scalar_JF}
\ddot{\varphi}=-3H\dot{\varphi}+4\phi_{0}\left(\frac{\Lambda}{3}-\frac{\tilde{\Lambda}}{3}\left(\frac{\varphi}{\phi_{0}}-1\right)\right)\,,
\end{equation}
where an over-dot denotes differentiation with respect to cosmological time $t$.

The dynamical equation for the scalar file \eqref{eq:scalar_JF}, together with the acceleration equation \eqref{eq:accel_JF}, subject to the energy conservation condition \eqref{eq:en_JF}, completely determine the dynamical behaviour of the model. Various dimensionless dynamical variables \cite{Hrycyna:2010yv,Hrycyna:2015eta,Hrycyna:2020jmw,Kerachian:2019tar,Jarv:2021qpp} can be introduced to parametrise the phase space and investigate the dynamics of the model.

In section \ref{sec:dyn_jf}, we present a complete dynamical system analysis of the model, where we find all the asymptotic states and determine their character and physical interpretation.

\subsection{Einstein frame}

When investigating modified and extended theories of gravity, one usually employs a conformal transformation procedure to achieve mathematical simplification of the theory \cite{Stelle:1977ry, Whitt:1984pd}. However, we should be aware of the important ontological consequences of this technique, particularly as it results in matter generation through a mathematical procedure.

The conformal transformation of the metric tensor
$$
\tilde{g}_{\mu\nu}=\Omega^{2}g_{\mu\nu}\,,
$$
with the conformal factor
\begin{equation}
\label{eq:confac}
\Omega^{2}=\frac{\varphi}{\phi_{0}}\,,
\end{equation}
transforms the action integral \eqref{eq:action_jf} of the Jordan frame into the Einstein frame:
\begin{equation}
\label{eq:ac_tot_ef}
\tilde{S} = \tilde{S}_{g}+16\pi \tilde{S}_{\varphi}\,,
\end{equation}
where the gravitational part is given by
\begin{equation}
\label{eq:ac_g_ef}
\tilde{S}_{g}=\phi_{0}\int\ud^{4}x\sqrt{-\tilde{g}}\,\tilde{R}\,,
\end{equation}
and the substantial matter part is in form of the scalar field with the following action integral:
\begin{equation}
\label{eq:ac_phi_ef}
\tilde{S}_{\varphi}=-\frac{1}{16\pi}\int\ud^{4}x\sqrt{-\tilde{g}}\left(\frac{\omega(\varphi)}{\varphi}\tilde{g}^{\mu\nu}\nabla_{\mu}\varphi\,\nabla_{\nu}\varphi+2\tilde{V}(\varphi)\right)
\end{equation}
where
$$
\omega(\varphi)=\frac{3}{2}\frac{\phi_{0}}{\varphi}\,, \quad 
\tilde{V}(\varphi) = \phi_{0}\frac{\phi_{0}^{2}}{\varphi^{2}}\left(\Lambda+\tilde{\Lambda}\bigg(\frac{\varphi}{\phi_{0}}-1\bigg)^{2}\right)\,,
$$
and
$$
\tilde{\Lambda}=\frac{\phi_{0}}{8c_{1}}\,,
$$
is the cosmological constant in the Einstein frame.

We can use the following scalar field redefinition:
$$
\frac{\varphi}{\phi_{0}}=\exp{\left(\sqrt{\frac{16\pi}{3\phi_{0}}}\tilde{\phi}\right)}\,,
$$
to work with a canonically normalised, minimally coupled scalar field cosmology with the following action integral:
\begin{equation}
\label{eq:a_norm}
S^{(E)} = S_{g}^{(E)} + 16\pi S_{\tilde{\phi}}^{(E)}\,.
\end{equation}
The gravitational part of the theory can be presented as
\begin{equation}
S_{g}^{(E)} = \phi_{0}\int\ud^{4}x\sqrt{-\tilde{g}}\left(\tilde{R}-2\tilde{\Lambda}\right)\,,
\end{equation}
while the normalised scalar field action is
\begin{equation}
S_{\tilde{\phi}}^{(E)} = -\frac{1}{2}\int\ud^{4}x\sqrt{-\tilde{g}}
\left(\tilde{g}^{\mu\nu}\nabla_{\mu}\tilde{\phi}\,\nabla_{\nu}\tilde{\phi}+2U(\tilde{\phi})\right)\,,
\end{equation}
where the scalar field potential function is now given by
$$
U(\tilde{\phi}) = \frac{\phi_{0}\tilde{\Lambda}}{4\pi}
\left(\frac{\Lambda+\tilde{\Lambda}}{2\tilde{\Lambda}}\exp{\left(-\sqrt{\frac{16\pi}{3\phi_{0}}}\tilde{\phi}\right)}-1\right)
\exp{\left(-\sqrt{\frac{16\pi}{3\phi_{0}}}\tilde{\phi}\right)}\,.
$$
Within the dynamical system approach used in this paper, we can easily show that the physical and dynamical description of the model given by the action integral \eqref{eq:ac_tot_ef} are completely equivalent to those given by the action \eqref{eq:a_norm}. To facilitate a straightforward dynamical and physical comparison between the frames, we will work with the original, non-normalised scalar field and the action integral for the theory \eqref{eq:ac_tot_ef}.

Variation of the action integral with respect to the metric tensor,	 $\delta\tilde{S}/\delta\tilde{g}^{\mu\nu}=0$, gives the following field equations: 
\begin{equation}
\tilde{R}_{\mu\nu}-\frac{1}{2}\tilde{g}_{\mu\nu}\tilde{R}=\frac{8\pi}{\phi_{0}} \tilde{T}_{\mu\nu}^{(\varphi)}
\end{equation}
where the energy-momentum tensor for the scalar field is
\begin{equation}
\tilde{T}_{\mu\nu}^{(\varphi)} = \frac{1}{8\pi}\frac{\omega(\varphi)}{\varphi}\nabla_{\mu}\varphi\,\nabla_{\nu}\varphi-\frac{1}{16\pi}\frac{\omega(\varphi)}{\varphi}\tilde{g}_{\mu\nu}\tilde{g}^{\alpha\beta}\nabla_{\alpha}\varphi\,\nabla_{\beta}\varphi - \frac{1}{8\pi}\tilde{V}(\varphi)\tilde{g}_{\mu\nu}\,.
\end{equation}
The trace of the field equations is
\begin{equation}
\tilde{R}=-\frac{8\pi}{\phi_{0}}\tilde{T}^{(\varphi)}\,,
\end{equation}
where the trace of the energy-momentum tensor of the scalar field is
\begin{equation}
\tilde{T}^{(\varphi)} = - \frac{1}{8\pi}\frac{\omega(\varphi)}{\varphi}\tilde{g}^{\mu\nu}\nabla_{\mu}\varphi\,\nabla_{\nu}\varphi-\frac{1}{8\pi}4\tilde{V}(\varphi)\,.
\end{equation}

Variation of the scalar field action integral, $\delta\tilde{S}_{\varphi}/\delta\varphi$=0, gives the equation of motion for the scalar field:
\begin{equation}
2\frac{\omega(\varphi)}{\varphi}\tilde{\Box}\varphi+\left(\frac{\omega'(\varphi)}{\varphi}-\frac{\omega(\varphi)}{\varphi^{2}}\right)\tilde{g}^{\mu\nu}\nabla_{\mu}\varphi\,\nabla_{\nu}\varphi-2\tilde{V}'(\varphi)=0\,.
\end{equation}

Working with the Einstein frame spatially flat FRW metric
$$
\ud\tilde{s}^{2} = -\ud\tilde{t}^{2} + \tilde{a}^{2}\left(\tilde{t}\right)\Big(\ud x^{2}+\ud y^{2}+\ud z^{2}\Big)\,,
$$
we obtain the energy conservation condition:
\begin{equation}
\label{eq:en_EF}
\tilde{H}^{2} = \frac{1}{4}\frac{\dot{\varphi}^{2}}{\varphi^{2}} + \frac{\phi_{0}^{2}}{\varphi^{2}}\left(\frac{\Lambda}{3}+\frac{\tilde{\Lambda}}{3}\left(\frac{\varphi}{\phi_{0}}-1\right)^{2}\right)\,,
\end{equation}
the acceleration equation from the trace of the field equations:
\begin{equation}
\label{eq:accel_EF}
\dot{\tilde{H}}=-\frac{3}{4}\frac{\dot{\varphi}^{2}}{\varphi^{2}}
\end{equation}
and the equation of motion for the scalar field:
\begin{equation}
\label{eq:scalar_EF}
\ddot{\varphi}=-3\tilde{H}\dot{\varphi}+\frac{\dot{\varphi}^{2}}{\varphi}+ 4 \frac{\phi_{0}^{2}}{\varphi}\left(\frac{\Lambda}{3}-\frac{\tilde{\Lambda}}{3}\left(\frac{\varphi}{\phi_{0}}-1\right)\right)\,,
\end{equation}
where an over-dot denotes differentiation with respect to the Einstein frame cosmological time $\tilde{t}$.

In section \ref{sec:dyn_ef} we present a dynamical system analysis of the investigated model. 

\section{Dynamical system analysis}

\subsection{Jordan frame}
\label{sec:dyn_jf}

To parametrise the phase space and investigate the dynamical behaviour of the model, we introduce the following dimensionless dynamical variables \cite{Hrycyna:2010yv,Hrycyna:2015eta,Hrycyna:2020jmw,Hrycyna:2021yad}
\begin{equation}
x=\frac{\dot{\varphi}}{H\varphi}\,,\quad z=\frac{\varphi}{\phi_{0}}\,.
\end{equation}
Then, the energy conservation condition \eqref{eq:en_JF}, normalised to the present value of the Hubble constant $H_{0}$, can be written in the following form:
\begin{equation}
\label{eq:en_j}
\frac{H^{2}}{H_{0}^{2}}=\Omega_{\tilde{\Lambda},0}\frac{\lambda+(z-1)^{2}}{z(1+x)}=\frac{\Omega_{\Lambda,0}}{\lambda}\frac{\lambda+(z-1)^{2}}{z(1+x)}\,,
\end{equation}
where $\Omega_{\Lambda,0}=\frac{\Lambda}{3 H_{0}^{2}}$ is the present value of the energy density associated with the Jordan frame cosmological constant, while $\Omega_{\tilde{\Lambda},0}=\frac{\tilde{\Lambda}}{3 H_{0}^{2}}$ is the present value of the energy density associated with the Einstein frame cosmological constant $\tilde{\Lambda}$. The parameter $\lambda$ is defined as
$$\lambda=\frac{\Omega_{\Lambda,0}}{\Omega_{\tilde{\Lambda},0}}=\frac{\Lambda}{\tilde{\Lambda}}=8c_{1}\frac{\Lambda}{\phi_{0}}\,.$$

The acceleration equation \eqref{eq:accel_JF} expressed in terms of the dimensionless dynamical variables is:
\begin{equation}
\label{eq:accel_j}
\frac{\dot{H}}{H^{2}}=-2+2(x+1)\frac{z(z-1)}{\lambda+(z-1)^{2}}\,.
\end{equation}
Finally, using the equation of motion for the scalar field \eqref{eq:scalar_JF}, we obtain the following dynamical system:
\begin{equation}
\label{eq:dyn_sys_j}
\begin{split}
\frac{\ud x}{\ud \ln{a}} &= -x(x+1)-x\left(\frac{\dot{H}}{H^{2}}+2\right)-4(1+x)\frac{z-(1+\lambda)}{\lambda+(z-1)^{2}}\,,\\
\frac{\ud z}{\ud \ln{a}} &= x z\,,
\end{split}
\end{equation}
which completely describes the dynamics of the model under investigation.

In the dynamical systems approach to cosmological models, we are interested in asymptotic states of the dynamics, which can correspond to various epochs in the evolution of the universe. Asymptotic states are points in the phase space where the right-hand sides of the dynamical equations vanish.

First, we find two saddle type critical points located at $(x^{*}=-1, z^{*}=0)$ and $(x^{*}=4, z^{*}=0)$. The acceleration equation \eqref{eq:accel_j}, evaluated at these points, gives
$$
\frac{\dot{H}}{H^{2}}\bigg|^{*} = -2\,.
$$
Physically, these states correspond to a transient radiation-like expansion of the universe.

The most interesting critical point is located at $(x^{*}=0, z^{*}=1+\lambda)$, with the energy conservation condition and the acceleration equation calculated at this point:
$$
\frac{H^{2}}{H_{0}^{2}}\bigg|^{*} = \Omega_{\Lambda,0}\,, \quad 
\frac{\dot{H}}{H^{2}}\bigg|^{*}=0\,,
$$
which corresponds to the de Sitter state. The eigenvalues of the linearisation matrix are 
$$
l_{1} = -\frac{3}{2} - \frac{1}{2}\sqrt{9-\frac{16}{\lambda}}\,, \quad
l_{2} = -\frac{3}{2} + \frac{1}{2}\sqrt{9-\frac{16}{\lambda}}\,,
$$
and we determine the character of the critical point as follows: for $\lambda<0$ : $l_{1}<0$, $l_{2}>0$ -- a saddle; for $0<\lambda<\frac{16}{9}$ : $Re[l_{1}]<0$, $Re[l_{2}]<0$ -- a stable focus (a sink); for $\lambda\ge\frac{16}{9}$ : $l_{1}<0$, $l_{2}<0$ -- a stable node. We conclude that the de Sitter state under consideration corresponds to an asymptotically stable critical point for $\lambda>0$. Since $\lambda=\Lambda/\tilde{\Lambda}$, we deduce that, to ensure stability of the de Sitter state, both the Jordan frame cosmological constant $\Lambda$ and the Einstein frame cosmological constant $\tilde{\Lambda}$ must be positive.

The linearised solutions in the vicinity of the de Sitter state are given by
\begin{equation}
\label{eq:linsol_j}
\begin{split}
x(a) & = x^{*} + l_{1}\frac{(1+\lambda)\Delta x-l_{2}\Delta z}{(1+\lambda)(l_{1}-l_{2})}\left(\frac{a}{a^{(i)}}\right)^{l_{1}} - l_{2}\frac{(1+\lambda)\Delta x-l_{1}\Delta z}{(1+\lambda)(l_{1}-l_{2})}\left(\frac{a}{a^{(i)}}\right)^{l_{2}}\,,\\
z(a) & = z^{*} + \frac{(1+\lambda)\Delta x - l_{2}\Delta z}{l_{1}-l_{2}}\left(\frac{a}{a^{(i)}}\right)^{l_{1}} - \frac{(1+\lambda)\Delta x - l_{1}\Delta z}{l_{1}-l_{2}}\left(\frac{a}{a^{(i)}}\right)^{l_{2}}\,.
\end{split}
\end{equation}
where $\Delta x = x^{(i)} - x^{*}$ and $\Delta z = z^{(i)}-z^{*}$ are the initial conditions in the phase space, and $a^{(i)}$ is the initial value of the scale factor.

The linearised solutions can be used to find the approximate behaviour of the Hubble function \eqref{eq:en_j} in the vicinity of the de Sitter state. Up to quadratic terms in the initial conditions $\Delta x$ and $\Delta z$, we find the following:
\begin{equation}
\begin{split}
\frac{H^{2}}{H_{0}^{2}} &\approx  \Omega_{\Lambda,0} \\
& \quad + \frac{\Omega_{\Lambda,0}}{1+\lambda}\Bigg((1-l_{1})\frac{(1+\lambda)\Delta x- l_{2}\Delta z}{l_{1}-l_{2}}\left(\frac{a}{a^{(i)}}\right)^{l_{1}}\\
& \qquad \qquad \hspace{1mm}-(1-l_{2})\frac{(1+\lambda)\Delta x-l_{1}\Delta z}{l_{1}-l_{2}}\left(\frac{a}{a^{(i)}}\right)^{l_{2}}\Bigg) \\
& \quad + \frac{\Omega_{\Lambda,0}}{\lambda(1+\lambda)^{2}}
\Bigg((1-\lambda l_{1}+\lambda l_{1}^{2})\left(\frac{(1+\lambda)\Delta x- l_{2}\Delta z}{l_{1}-l_{2}}\right)^{2}\left(\frac{a}{a^{(i)}}\right)^{2l_{1}}\\
& \qquad \qquad \qquad + 
(1-\lambda l_{2}+\lambda l_{2}^{2})\left(\frac{(1+\lambda)\Delta x- l_{1}\Delta z}{l_{1}-l_{2}}\right)^{2}\left(\frac{a}{a^{(i)}}\right)^{2l_{2}}\\
& \qquad \qquad \qquad -(10+3\lambda)\frac{((1+\lambda)\Delta x- l_{1}\Delta z)((1+\lambda)\Delta x- l_{2}\Delta z)}{(l_{1}-l_{2})^{2}}\left(\frac{a}{a^{(i)}}\right)^{-3}
\Bigg)\,,
\end{split}
\end{equation}
where the last term can be naturally interpreted as an effective dark matter component in the model,
$$
\Omega_{dm} = \Omega_{dm,i}\left(\frac{a}{a^{(i)}}\right)^{-3},
$$
and the initial value of the energy density of the effective dark matter can be expressed as
\begin{equation}
\label{eq:dm_JF}
\Omega_{dm,i}=-\Omega_{\Lambda,0}\frac{10+3\lambda}{\lambda(1+\lambda)^{2}\left(9-\frac{16}{\lambda}\right)}\left(\left((1+\lambda)\Delta x+\frac{3}{2}\Delta z\right)^{2}-\frac{1}{4}\left(9-\frac{16}{\lambda}\right)\Delta z^{2}\right)\,.
\end{equation}

This term has to be positive in order to effectively imitate a dark matter component in the theory. For $\Omega_{\Lambda,0}>0$ and $\lambda>\frac{16}{9}$, we can always find initial conditions in the phase space where $\Omega_{dm,i}>0$, while for $\Omega_{\Lambda,0}>0$ and $0<\lambda<\frac{16}{9}$, the sign of the term does not depend on the initial conditions.
We must remember that this effect is a result of the dynamical behaviour of the model.

Next, we proceed to investigate the dynamical behaviour of the model at infinite values of the scalar field $\varphi$. By introducing the dynamical variable $v=\frac{1}{z}$, we obtain the following dynamical system:
\begin{equation}
\label{eq:dyn_sys_j_proj}
\begin{split}
\frac{\ud x}{\ud \ln{a}} &= -(x+1)\left(x+2x\frac{1-v}{\lambda v^{2}+(1-v)^{2}}+4v\frac{1-(1+\lambda)v}{\lambda v^{2}+(1-v)^{2}}\right)\,,\\
\frac{\ud v}{\ud \ln{a}} &= -x v\,,
\end{split}
\end{equation}
and, in the same way, we obtain the energy conservation condition \eqref{eq:en_j} and the acceleration equation \eqref{eq:accel_j}. 

\begin{figure}
\centering
\includegraphics[scale=0.675]{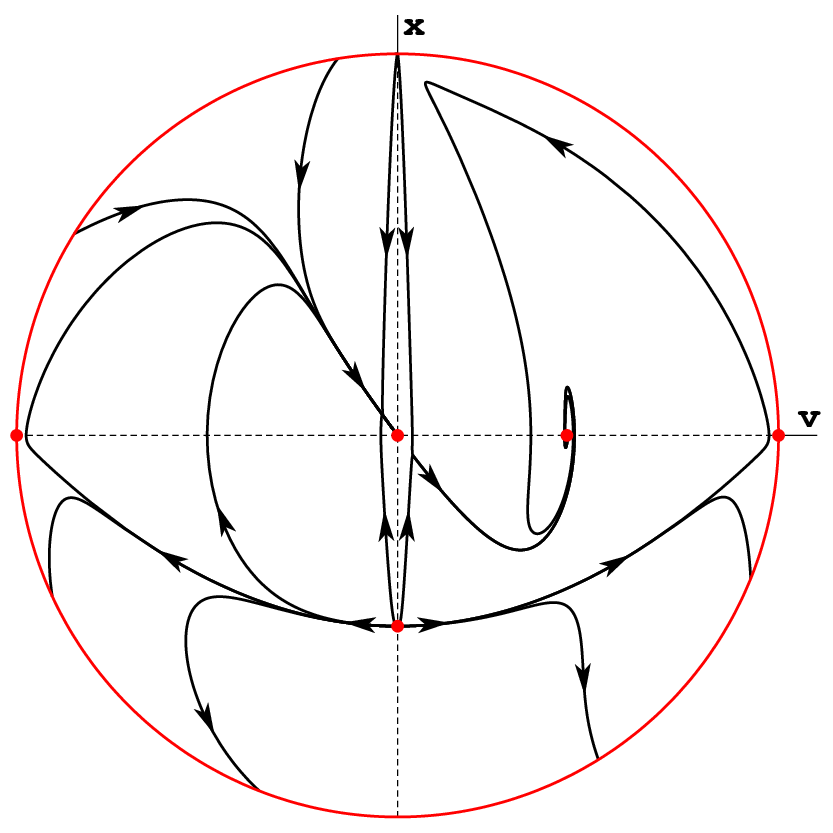}
\caption{The phase space diagram of the dynamical system \eqref{eq:dyn_sys_j_proj} in the Jordan frame with $\lambda=\frac{1}{4}$. The invariant centre manifold and a saddle-node degenerated critical point at the centre are clearly visible.}
\label{fig:1}
\end{figure}

Using the standard procedure, we search for asymptotic states located at $v^{*}=0$. We find two critical points corresponding to infinite values of the scalar field  $\phi$. The first critical point, located at $(x^{*}=-1, v^{*}=0)$ is an unstable node with the following linearised solutions in the vicinity of this state:
\begin{equation}
\begin{split}
x(a) & = -1 +\Delta x \left(\frac{a}{a^{(i)}}\right)^{3}\,,\\
v(a) & = \Delta v \left(\frac{a}{a^{(i)}}\right)\,,
\end{split}
\end{equation}
where $\Delta x$, $\Delta z$ are the initial conditions, and $a^{(i)}$ is the initial value of the scale factor. The energy conservation condition \eqref{eq:en_j} and the acceleration equation \eqref{eq:accel_j} calculated at this state give
$$
\frac{H^{2}}{H_{0}^{2}}\bigg|^{*}=\infty\,, \quad \frac{\dot{H}}{H^{2}}\bigg|^{*} = -2\,,
$$
which suggest that the effective equation of state parameter is $w_{\text{eff}}=\frac{1}{3}$ and it corresponds to a radiation-like expansion of the universe.

The second asymptotic state, located at $(x^{*}=0, v^{*}=0)$, is very interesting from both a mathematical and physical point of view. The eigenvalues of the linearisation matrix calculated at this point are
$$
l_{1}=-3\,, \quad l_{2}=0\,,
$$
and, with one of the eigenvalues vanishing, we have a degenerate non-hyperbolic critical point. The stability analysis requires the use of the centre manifold theorem \cite{Perko:book,Wiggins:book,Hrycyna:2010yv}. 

According to this theorem, we can easily find the following equation for an invariant manifold:
$$
x = -\frac{4}{3}v+ f(v)\,,
$$
where the the function $f(v)$ is an arbitrary polynomial:
$$
f(v) = a_{2}v^{2} + a_{3}v^{3} + a_{4} v^{4} + a_{5} v^{5} \ldots\,,
$$
and this manifold is called the centre manifold. Using the centre manifold theorem, we can find the constants $a_{i}$ to an arbitrary order:
\begin{equation}
\begin{split}
a_{2} & = \frac{4}{27}(1+9\lambda)\,,\\
a_{3} & = \frac{4}{81}(1+9\lambda)\,,\\
a_{4} & = \frac{4}{2187}(1+9\lambda)(29+45\lambda)\,,\\
a_{5} & = \frac{4}{19683}(1+9\lambda)(23-9\lambda)\,,\\
\vdots\,
\end{split}
\end{equation} 
In this way, we can find the equation for the invariant centre manifold to arbitrary precision. Finally, we conclude that the asymptotic state under consideration results from a saddle-node bifurcation and can be split by adding an arbitrarily small distortion to the model \cite{Perko:book, Wiggins:book}.

The energy conservation condition \eqref{eq:en_j} and the acceleration equation \eqref{eq:accel_j} calculated at the asymptotic state under considerations are
$$
\frac{H^{2}}{H_{0}^{2}}\bigg|^{*}=\infty\,, \quad \frac{\dot{H}}{H^{2}}\bigg|^{*} = 0\,, \quad \frac{\dot{H}}{H_{0}^{2}}\bigg|^{*}= -\frac{2}{3\lambda}\Omega_{\Lambda,0}\,.
$$
Then, using the invariant centre manifold, we find that,as we approach the critical point, the scale factor
$$
a \to 0\,, \quad \dot{a} \to 0\,, \quad \ddot{a} \to 0\,,
$$
tends to zero faster than $\dot{a}$ and $\ddot{a}$.

In Figure~\ref{fig:1}, we present the phase space diagram of the dynamical system \eqref{eq:dyn_sys_j_proj}, compactified with a circle at infinity, without enforcing any physical interpretation of the dynamics. Direction of arrows on the phase space curves indicates growing of the scale factor. We can readily observe the shape of the invariant centre manifold connecting the critical point at the centre of the phase space $(x^{*}=0, v^{*}=0)$ with the point at $(x^{*}=0, v^{*}=\frac{1}{1+\lambda})$. An interesting property of the dynamics is the existence of a critical circle at infinity in the phase space.

\subsection{Einstein frame}
\label{sec:dyn_ef}

Following the prescription from the previous section on the Jordan frame dynamics, we introduce the following dimensionless dynamical variables:
\begin{equation}
\tilde{x}=\frac{\dot{\varphi}}{\tilde{H}\varphi}\,, \quad z=\frac{\varphi}{\phi_{0}}
\end{equation}
where an over-dot now denotes differentiation with respect to the Einstein frame cosmological time $\tilde{t}$, and $\tilde{H}$ is the Einstein frame Hubble function $\tilde{H}=\frac{\ud\ln{\tilde{a}}}{\ud\tilde{t}}$.

The energy conservation condition \eqref{eq:en_EF}, normalised to the present value of the Einstein frame Hubble constant, is 
\begin{equation}
\label{eq:en_e}
\frac{\tilde{H}^{2}}{\tilde{H}_{0}^{2}}=
\frac{\tilde{\Omega}_{\Lambda,0}}{\lambda}\frac{\lambda+(z-1)^{2}}{z^{2}\left(1-\frac{1}{4}\tilde{x}^{2}\right)}=\tilde{\Omega}_{\tilde{\Lambda},0}\frac{\lambda+(z-1)^{2}}{z^{2}\left(1-\frac{1}{4}\tilde{x}^{2}\right)}\,,
\end{equation}
where $\tilde{\Omega}_{\Lambda,0}=\frac{\Lambda}{3 \tilde{H}_{0}^{2}}$ is the present value of the energy density resulting from the Jordan frame cosmological constant, and $\tilde{\Omega}_{\tilde{\Lambda},0}=\frac{\tilde{\Lambda}}{3 \tilde{H}_{0}^{2}}$ is the present value of the energy density of the Einstein frame cosmological constant. Their ratio is 
$$\lambda=\frac{\tilde{\Omega}_{\Lambda,0}}{\tilde{\Omega}_{\tilde{\Lambda},0}}=\frac{\Lambda}{\tilde{\Lambda}}=8c_{1}\frac{\Lambda}{\phi_{0}}\,.$$
The acceleration equation \eqref{eq:accel_EF} expressed in the dimensionless variables is
\begin{equation}
\label{eq:accel_e}
\frac{\dot{\tilde{H}}}{\tilde{H}^{2}}=-\frac{3}{4}\tilde{x}^{2}\,.
\end{equation}
Finally, using the equation of motion for the scalar field \eqref{eq:scalar_EF}, we find the following dynamical system, which completely describes the model under investigation:
\begin{equation}
\label{eq:dyn_sys_e}
\begin{split}
\frac{\ud \tilde{x}}{\ud \ln{\tilde{a}}} &= -3\left(1-\frac{1}{4}\tilde{x}^{2}\right)
\left(\tilde{x}-\frac{4}{3}\frac{1+\lambda-z}{\lambda+(z-1)^{2}}\right)\,,\\
\frac{\ud z}{\ud \ln{\tilde{a}}} &= \tilde{x} z\,,
\end{split}
\end{equation}
where the ``time parameter'' along phase curves is the Einstein frame scale factor $\tilde{a}$.

Solving the right-hand side of the dynamical system, we find four critical points in the phase space. One, located at $(\tilde{x}^{*}=\frac{4}{3}, z^{*}=0)$, is a saddle-type critical point with the effective equation of state parameter $w_{\text{eff}}=-\frac{1}{9}$. The next two, with effective equation of state parameter $w_{\text{eff}}=1$ corresponding to Zeldovich stiff matter \cite{Zeldovich:1962,Zeldovich:1972zz}. One, located at $(\tilde{x}^{*}=+2, z^{*}=0)$ is an unstable node, while the second, located at $(\tilde{x}^{*}=-2, z^{*}=0)$, is a saddle.

The final critical point, located at $(\tilde{x}^{*}=0, z^{*}=1+\lambda)$, with the energy conservation condition and the acceleration equation 
$$
\frac{\tilde{H}^{2}}{\tilde{H}_{0}^{2}}\bigg|^{*} =\frac{\tilde{\Omega}_{\Lambda,0}}{1+\lambda}\,, \quad 
\frac{\dot{\tilde{H}}}{\tilde{H}^{2}}\bigg|^{*}=0\,,
$$
corresponds to the de Sitter state. The eigenvalues of the linearisation matrix are 
$$
l_{1} = -\frac{3}{2} - \frac{1}{2}\sqrt{9-\frac{16}{\lambda}}\,, \quad
l_{2} = -\frac{3}{2} + \frac{1}{2}\sqrt{9-\frac{16}{\lambda}}\,,
$$
and we determine the character of the critical point as follows: for $\lambda<0$ : $l_{1}<0$, $l_{2}>0$ -- a saddle; for $0<\lambda<\frac{16}{9}$ : $Re[l_{1}]<0$, $Re[l_{2}]<0$ -- a stable focus (a sink); for $\lambda\ge\frac{16}{9}$ : $l_{1}<0$, $l_{2}<0$ -- a stable node.

The linearised solutions near the de Sitter state are
\begin{equation}
\begin{split}
\tilde{x}(\tilde{a}) & = x^{*}+ 
l_{1}\frac{(1+\lambda)l_{1}\Delta\tilde{x}-l_{2}\Delta z}
{(1+\lambda)(l_{1}-l_{2})}\left(\frac{\tilde{a}}{\tilde{a}^{(i)}}\right)^{l_{1}}
- l_{2}\frac{(1+\lambda)l_{2}\Delta\tilde{x}-l_{1}\Delta z}
{(1+\lambda)(l_{1}-l_{2})}\left(\frac{\tilde{a}}{\tilde{a}^{(i)}}\right)^{l_{2}} \,,\\
z(\tilde{a}) & = z^{*} + \frac{(1+\lambda)\Delta\tilde{x} - l_{2}\Delta z}{l_{1}-l_{2}}
\left(\frac{\tilde{a}}{\tilde{a}^{(i)}}\right)^{l_{1}}
-\frac{(1+\lambda)\Delta\tilde{x}-l_{1}\Delta z}{l_{1}-l_{2}}
\left(\frac{\tilde{a}}{\tilde{a}^{(i)}}\right)^{l_{2}}\,,
\end{split}
\end{equation}
where $\Delta\tilde{x}=\tilde{x}^{(i)}-\tilde{x}^{*}$ and $\Delta z = z^{(i)} - z^{*}$ are the initial conditions in the phase space, while $\tilde{a}^{(i)}$ is the initial value of the scale factor in the Einstein frame.

We note that the stability conditions and the linearised solutions have exactly the same form as in the Jordan frame case.

Using the linearised solution to expand the Hubble function \eqref{eq:en_e} up to quadratic terms in initial conditions $\Delta\tilde{x}$ and $\Delta z$, we find the following approximation:
\begin{equation}
\begin{split}
\frac{\tilde{H}^{2}}{\tilde{H}_{0}^{2}} &\approx  \frac{\tilde{\Omega}_{\Lambda,0}}{1+\lambda} \\
& \quad + \frac{\tilde{\Omega}_{\Lambda,0}}{4\lambda(1+\lambda)^{3}}
\Bigg((4+\lambda l_{1}^{2})\left(\frac{(1+\lambda)\Delta\tilde{x}- l_{2}\Delta z}{l_{1}-l_{2}}\right)^{2}\left(\frac{\tilde{a}}{\tilde{a}^{(i)}}\right)^{2l_{1}}\\
& \quad \hspace{23.2mm}+ 
(4+\lambda l_{2}^{2})\left(\frac{(1+\lambda)\Delta\tilde{x}- l_{1}\Delta z}{l_{1}-l_{2}}\right)^{2}\left(\frac{\tilde{a}}{\tilde{a}^{(i)}}\right)^{2l_{2}}\\
& \quad \hspace{23.2mm} -16\frac{((1+\lambda)\Delta\tilde{x}- l_{1}\Delta z)((1+\lambda)\Delta\tilde{x}- l_{2}\Delta z)}{(l_{1}-l_{2})^{2}}\left(\frac{\tilde{a}}{\tilde{a}^{(i)}}\right)^{-3}
\Bigg)\,,
\end{split}
\end{equation}
where the last term in this equation can be naturally interpreted as an effective dark matter component of the model:
$$
\tilde{\Omega}_{dm}=\tilde{\Omega}_{dm,i}\left(\frac{\tilde{a}}{\tilde{a}^{(i)}}\right)^{-3}
$$
and the initial value of the energy density is
\begin{equation}
\tilde{\Omega}_{dm,i} = -4\frac{\tilde{\Omega}_{\Lambda,0}}{\lambda(1+\lambda)^{3}\left(9-\frac{16}{\lambda}\right)}\left(\left((1+\lambda)\Delta\tilde{x}+\frac{3}{2}\Delta z\right)^{2}-\frac{1}{4}\left(9-\frac{16}{\lambda}\right)\Delta z^{2}\right)\,.
\end{equation}

The conditions for the positivity of this term are exactly the same as in the Jordan frame case. For an asymptotically stable de Sitter state, we have $\lambda>0$. Then, for $0<\lambda<\frac{16}{9}$, the sign of $\tilde{\Omega}_{dm,i}$ does not depend on the initial conditions $\Delta\tilde{x}$ and $\Delta z$. For $\lambda>\frac{16}{9}$, we can always find regions of the phase space where $\tilde{\Omega}_{dm,i}$ is positive. Again, we must remember that this effective dark matter term is the result of the dynamical behaviour of the model.

\begin{figure}
\centering
\includegraphics[scale=0.675]{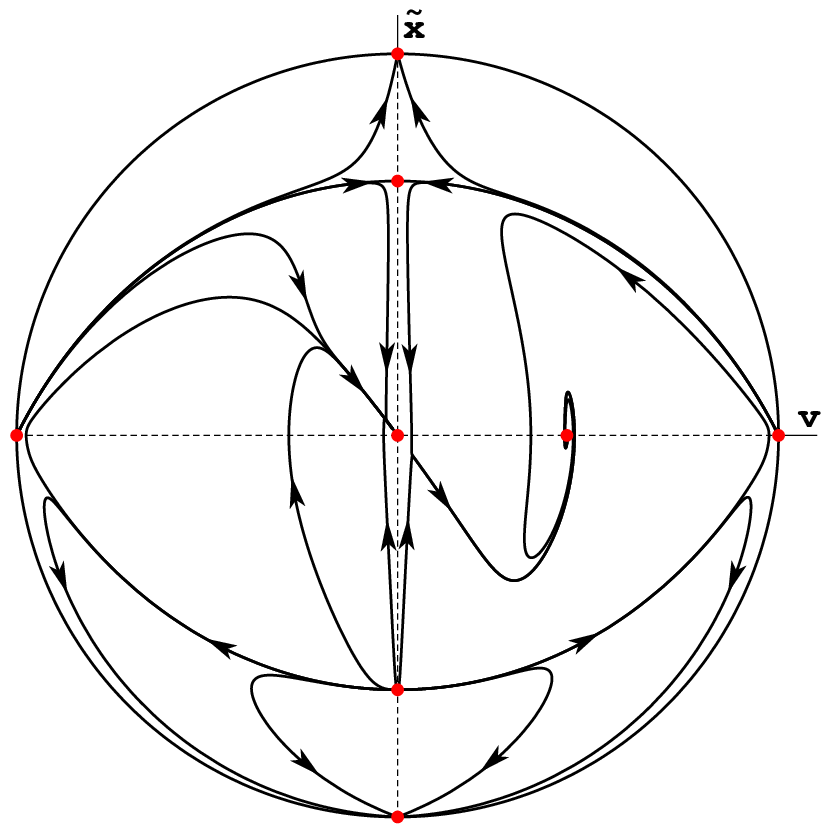}
\caption{The phase space diagram of the dynamical system \eqref{eq:dyn_sys_e_proj} in the Einstein frame with $\lambda=\frac{1}{4}$. The invariant centre manifold and a saddle-node degenerated critical point at the centre are clearly visible.}
\label{fig:2}
\end{figure}

We can now proceed to investigate dynamics of the model at infinite values of the scalar field $\varphi$ in the same manner as in the Jordan frame case. Introducing dynamical variable $v=\frac{1}{z}$, we find the following dynamical system:
\begin{equation}
\label{eq:dyn_sys_e_proj}
\begin{split}
\frac{\ud\tilde{x}}{\ud\ln{\tilde{a}}} &= -3\left(1-\frac{1}{4}\tilde{x}^{2}\right) \left(\tilde{x} + \frac{4}{3}v\frac{1-(1+\lambda)v}{\lambda v^{2}+(1-v)^{2}}\right)\,,\\
\frac{\ud v}{\ud\ln{\tilde{a}}} & = -\tilde{x} v\,,
\end{split}
\end{equation}
and in the same way we find the energy conservation condition \eqref{eq:en_e}.

Next, we look for asymptotic states located at infinite values of the scalar field $\varphi$. One of the states, located at $(\tilde{x}^{*}=2, v^{*}=0)$, is a saddle type critical point, and the second with the coordinates $(\tilde{x}^{*}=-2, v^{*}=0)$, is an unstable node. At both sates, the energy conservation condition is infinite, while the acceleration equation \eqref{eq:accel_e} is
$$
\frac{\dot{\tilde{H}}}{\tilde{H}^{2}}\bigg|^{*} = -3,
$$
with effective equation of state parameter $w_{\text{eff}}=1$, which corresponds to the Zeldovich stiff matter \cite{Zeldovich:1962,Zeldovich:1972zz}.

The most interesting critical point is located at the centre of the phase space $(\tilde{x}^{*}=0, v^{*}=0)$. The eigenvalues of the linearisation matrix calculated at this point are
$$
l_{1}=-3\,, \quad l_{2}=0\,,
$$
and, with one of the eigenvalues vanishing, we have a degenerate non-hyperbolic critical point. As in the Jordan frame case, the stability analysis requires the use of the centre manifold theorem \cite{Perko:book,Wiggins:book,Hrycyna:2010yv}. 

We can easily find the following equation for an invariant manifold: 
$$
\tilde{x} = -\frac{4}{3}v+ h(v)\,,
$$
where the the function $h(v)$ is an arbitrary polynomial:
$$
h(v) = a_{2}v^{2} + a_{3}v^{3} + a_{4} v^{4} + a_{5} v^{5} \ldots\,,
$$
and this manifold is called the centre manifold. Using the centre manifold theorem, we can find the constants  $a_{i}$ to an arbitrary order:
\begin{equation}
\begin{split}
a_{2} & = -\frac{4}{27}(5-9\lambda)\,,\\
a_{3} & = -\frac{4}{81}(7-45\lambda)\,,\\
a_{4} & = -\frac{4}{2187}(49-1494\lambda+81\lambda^{2})\,,\\
a_{5} & = -\frac{4}{19683}(-119-11970\lambda+6885\lambda^{2})\,,\\
\vdots\,
\end{split}
\end{equation} 
In this way, we can find the equation for the invariant centre manifold to arbitrary precision. Finally, we conclude that the asymptotic state under consideration results from a saddle-node bifurcation and can be split by adding an arbitrarily small distortion to the model  \cite{Perko:book, Wiggins:book}.

The energy conservation condition \eqref{eq:en_e} and the acceleration equation \eqref{eq:accel_e} calculated at this asymptotic state give
$$
\frac{\tilde{H}^{2}}{\tilde{H}_{0}^{2}}\bigg|^{*} = \frac{1}{\lambda}\tilde{\Omega}_{\Lambda,0} = \tilde{\Omega}_{\tilde{\Lambda},0}\,, \quad \frac{\dot{\tilde{H}}}{\tilde{H}^{2}}\bigg|^{*}=0\,,
$$
and with the constant energy conservation condition and the vanishing acceleration equation, we conclude that this critical point corresponds to the de Sitter state.

We must be aware that if the de Sitter state is reached asymptotically, the model might not be automatically non-singular. The scalar curvature invariants would be finite in this case, but a parallelly propagated singularity could be present \cite{Yoshida:2018ndv,Nomura:2021lzz}. Using the invariant centre manifold of the dynamics in the vicinity of the de Sitter state, we find
\begin{equation}
\begin{split}
\frac{\dot{\tilde{H}}}{\tilde{a}^{2}}\left(\frac{\tilde{a}^{(i)}}{\tilde{H}_{0}}\right)^{2} & = -\frac{3}{4}\frac{\tilde{\Omega}_{\Lambda,0}}{\lambda} \tilde{x}^{2}\frac{\lambda v^{2}+(1-v)^{2}}{1-\frac{1}{4}\tilde{x}^{2}} \left(\frac{\tilde{a}}{\tilde{a}^{(i)}}\right)^{-2} \\ 
& \approx -\frac{4}{3}\frac{\tilde{\Omega}_{\Lambda,0}}{\lambda} v^{2} \frac{\lambda v^{2}+(1-v)^{2}}{1-\frac{4}{9}v^{2}} \left(\frac{\tilde{a}}{\tilde{a}^{(i)}}\right)^{-2} \to -\infty \,,
\end{split}
\end{equation}
and this quantity diverges to infinity as $\tilde{a}\to0$ at the de Sitter state represented by the topological saddle-node bifurcation point.

Using the invariant centre manifold theorem, we have shown that there exists a zero-measure set of initial conditions that lead from an unstable to a stable de Sitter state. This suggests the possibility of transient cosmological evolution between different de Sitter phases.

In Figure~\ref{fig:2}, we present the complete phase space diagram of the dynamical system \eqref{eq:dyn_sys_e_proj}, compactified with a circle at infinity of the phase space. The direction of arrows on the phase space curves indicates growing of the Einstein frame scale factor $\tilde{a}$. One can easily notice presence of the invariant centre manifold connecting critical point at the centre of the phase space $(\tilde{x}^{*}=0, v^{*}=0)$ with the point at $(\tilde{x}^{*}=0, v^{*}=\frac{1}{1+\lambda})$. The saddle-node topological structure of the critical point $(\tilde{x}^{*}=0, v^{*}=0)$ is clear.

\section{Frames comparison and conclusions}

Now, we will investigate the correspondence between asymptotic states in the Jordan frame formulation of the theory and its conformally transformed Einstein frame counterpart. 
The conformal transformation between metric tensors is given by
$$
\tilde{g}_{\mu\nu} = \Omega^{2}g_{\mu\nu}\,,
$$ 
where $g_{\mu\nu}$ is the metric tensor in the Jordan frame and $\tilde{g}_{\mu\nu}$ is the metric tensor in the Einstein frame, with the conformal factor
$$
\Omega^{2} = \frac{\varphi}{\phi_{0}} > 0\,.
$$
We find that the scale factor and the cosmological time transform according to
$$
\tilde{a} = \Omega a\,, \qquad \ud\tilde{t} = \Omega \ud t\,.
$$

In dynamical system analysis, we have used phase space variables in the Einstein and Jordan frames: 
$$
\tilde{x}= \frac{\frac{\ud\varphi}{\ud\tilde{t}}}{\tilde{H}\varphi}\,,\qquad x = \frac{\frac{\ud\varphi}{\ud t}}{H\varphi}\,.
$$
 
\begin{figure}[h!]
\centering
\includegraphics[scale=0.5]{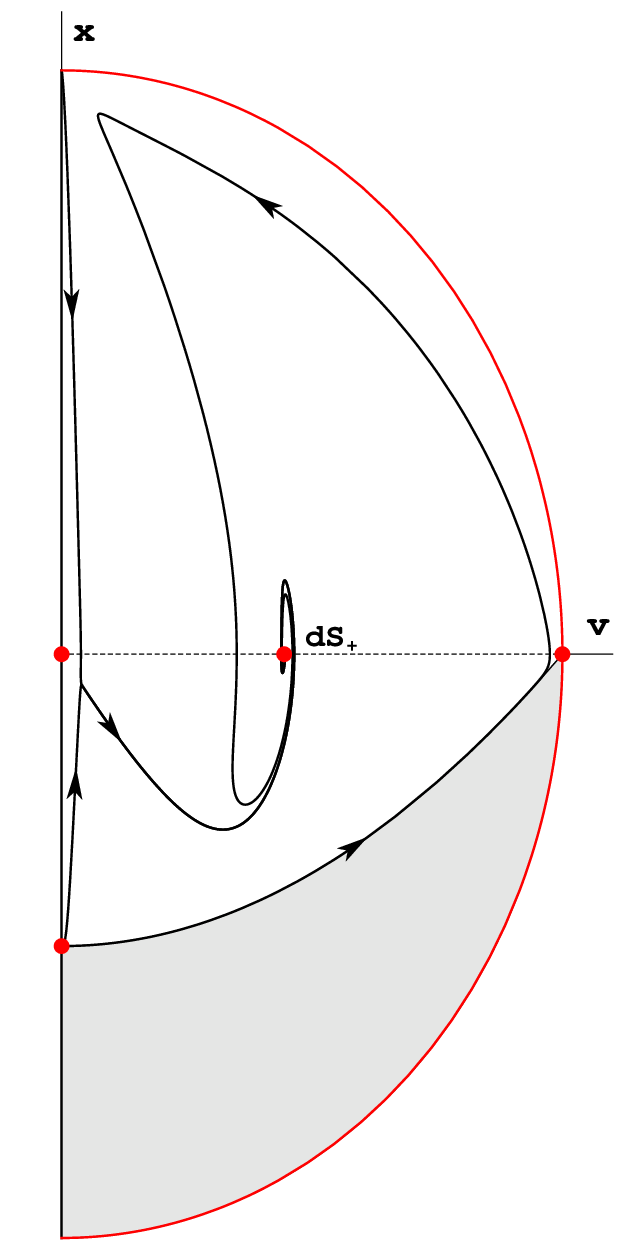}\hspace{2cm}
\includegraphics[scale=0.5]{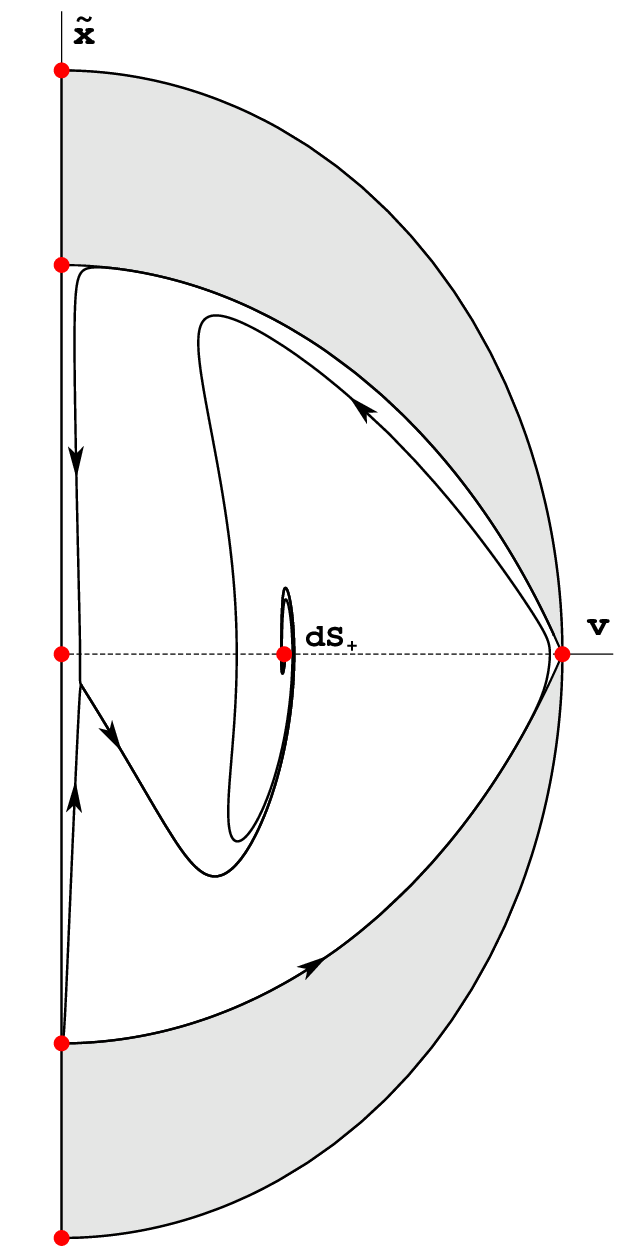}
\caption{Phase space diagrams in the Jordan frame and in the Einstein frame with $\lambda=\frac{1}{4}$. The shaded regions where $H^{2}<0$ or $\tilde{H}^{2}<0$ are unphysical. The invariant centre manifolds and the saddle-node degenerated critical points are clearly visible. The mathematical structures of the models are equivalent, but their physical content is different.}
\label{fig:3}
\end{figure}

Using the following transformations for the Hubble function between the frames,
$$
\tilde{H}=\frac{H}{\Omega}\bigg(1+\frac{1}{2}\frac{\ud\ln{\Omega^{2}}}{\ud\ln{a}}\bigg)\,,
$$
together with the conformal factor $\Omega^{2}=\frac{\varphi}{\phi_{0}}$, we find the following equations relating phase space variables in the Jordan frame and the Einstein frame: 
\begin{equation}
\label{eq:jve}
x =  \frac{\tilde{x}}{1-\frac{1}{2}\tilde{x}}\,, \quad
\tilde{x} =  \frac{x}{1+\frac{1}{2}x}\,.
\end{equation}
In Figure~\ref{fig:3}, we present the phase space diagrams in physical regions $\varphi>0$ for both frames. 

We begin our discussion of the correspondence between the frames with the Jordan frame critical point $(x^{*}=-1, v^{*}=0)$, which is an unstable node. The equation of state parameter calculated at this state is $w_{\text{eff}}=\frac{1}{3}$ which corresponds to radiation-like expansion of the universe. We have shown that this state is the single unstable critical point with an open and dense set of initial conditions giving rise to the beginning of the universe. The state under consideration is located at $\varphi^{*}=\infty$ where. in our analysis, $\varphi \propto R$, and from the original action integral of the theory \eqref{eq:act_orig}, we find that we arrive at an asymptotically conformally and scale-invariant theory. 

From the equation \eqref{eq:jve}, we find that the Jordan frame state $(x^{*}=-1, v^{*}=0)$ corresponds to the Einstein frame state $(\tilde{x}^{*}=-2, v^{*}=0)$. The Einstein frame physical state is described by Zeldovich stiff matter \cite{Zeldovich:1962,Zeldovich:1972zz}, with the equation of state parameter $w_{\text{eff}}=1$. We have found that in the Einstein frame analysis there are two such critical points in the form of an unstable node, which give rise to the beginning of a universe starting from the stiff matter state. This state is completely different from the Jordan frame state.

Next, we compare the centre of the phase space the critical points $(x^{*}=0, v^{*}=0)$ and $(\tilde{x}^{*}=0, v^{*}=0)$. In the Jordan frame, this state is described by the vanishing of the acceleration equation while the energy conservation condition diverges, and it corresponds to a singularity. In the Einstein frame, the corresponding state is of the de Sitter type but still plagued by a parallelly propagated singularity. The physical interpretation of both states is completely different.

The mathematical description of both states shows that they are in the form of degenerate non-hyperbolic critical points of a saddle-node bifurcation. For both states, we have found invariant centre manifolds with an arbitrary polynomial approximation. It is easy to show that both manifolds are invariant under the transformations~\eqref{eq:jve}. We conclude that the mathematical structure is preserved while the physical interpretation is changed.

The final critical points correspond to the de Sitter states. We can easily notice that the stability conditions and linearised solutions in the vicinity of the states are exactly the same in both frame. In the Jordan frame, the energy conservation condition for the asymptotic state is given by
$$
\frac{H^{2}}{H_{0}^{2}}\bigg|^{*} = \Omega_{\Lambda,0} = \frac{\Lambda}{3 H_{0}^{2}}\,,
$$
while in the Einstein frame it is
$$
\frac{\tilde{H}^{2}}{\tilde{H}_{0}^{2}}\bigg|^{*} = \frac{\tilde{\Omega}_{\Lambda,0}}{1+\lambda} = \frac{\frac{\Lambda}{3\tilde{H}_{0}^{2}}}{1+\lambda}\,,
$$
where $\lambda=\frac{\Lambda}{\tilde{\Lambda}}$ is the ratio between the Jordan frame cosmological constant and the Einstein frame cosmological constant. The energy density for the states in both frames is different unless measurements of the present value of the Hubble function depend on the frame.

In this paper, we investigated a theory of gravity derived from the truncated Sakharov action integral with a quadratic contribution of scalar curvature with the following Lagrangian $-2\phi_{0}\Lambda+\phi_{0}R+c_{1}R^{2}$. We have shown that when the simplest flat Friedmann--Robertson--Walker metric is employed the dynamics in both the Jordan frame and the conformally transformed Einstein frame can be reduced to two dimensional dynamical system. In both frames, we found stable de Sitter states where the expansion of the Hubble function naturally includes terms corresponding to an effective dark matter component. 

It is worth stressing that more general $f(R)$ theories of gravity represent viable cosmological solutions. In particular, cosmological models based on monomial scalar curvature terms $R^{n}$ in theories of gravity give rise to accelerated expansion as an attractor solution for a large set of initial conditions \cite{Carloni:2004kp}.

It is well known that the nature of solutions can change after a conformal transformation. While the mathematical structure of the dynamical system may be preserved, the physical interpretation of the solutions can be different in the Jordan and Einstein frames  \cite{Amendola:2006kh, Capozziello:2006dj}.

In our analysis, we have shown that the de Sitter states in both frames have the same stability conditions and linearised solutions. However, the energy densities associated with these states are different unless the measurements of the present value of the Hubble function are frame-dependent. We can conclude that while the qualitative behaviour of the solutions is preserved, the quantitative aspects can differ.

Moreover, the critical points corresponding to infinite values of the scalar field $\varphi$ exhibit different physical interpretations in the two frames. In the Jordan frame, the state $(x^{*}=-1, v^{*}=0)$ corresponds to a radiation-like expansion, while in the Einstein frame, the corresponding state $(\tilde{x}^{*}=-2, v^{*}=0)$ is described by Zeldovich stiff matter. This is why we have to carefully interpret the physical implications of solutions in different frames.

\bibliographystyle{elsarticle-num}
\bibliography{../bib/moje,../bib/darkenergy,../bib/quintessence,../bib/quartessence,../bib/astro,../bib/dynamics,../bib/standard,../bib/inflation,../bib/sm_nmc,../bib/singularities,../bib/JvsE}

\begin{thebibliography}{10}
\expandafter\ifx\csname url\endcsname\relax
  \def\url#1{\texttt{#1}}\fi
\expandafter\ifx\csname urlprefix\endcsname\relax\def\urlprefix{URL }\fi
\expandafter\ifx\csname href\endcsname\relax
  \def\href#1#2{#2} \def\path#1{#1}\fi

\bibitem{Sakharov:1967pk}
A.~D. Sakharov, {Vacuum quantum fluctuations in curved space and the theory of
  gravitation}, Gen.~Rel.~Grav. 32 (2000) 365--367, [Dokl.~Akad.~Nauk. SSSR
  {\bf 177}, 70--71 (1967); Sov.~Phys.~Dokl. {\bf 12}, 1040--1041, (1968);
  Sov.~Phys.~Usp. {\bf 34}, 394 (1991)].
\newblock \href {https://doi.org/10.1023/A:1001947813563}
  {\path{doi:10.1023/A:1001947813563}}.

\bibitem{Zeldovich:1967}
{\relax Ya}.~B. Zeldovich, {Interpretation of electrodynamics as a consequence
  of quantum theory}, JETP Lett. 6 (1967) 345--347, [Pisma Zh.~Eksp.~Teor.~Fiz.
  {\bf 6}, 922 (1967)].

\bibitem{Stelle:1977ry}
K.~S. Stelle, {Classical gravity with higher derivatives}, Gen.~Rel.~Grav. 9
  (1978) 353--371.
\newblock \href {https://doi.org/10.1007/BF00760427}
  {\path{doi:10.1007/BF00760427}}.

\bibitem{Adler:1983}
S.~L. Adler, {Einstein gravitation as a long wavelength effective field
  theory}, Phil.~Trans.~R.~Soc.~Lond.~A 310 (1983) 273--278.
\newblock \href {https://doi.org/10.1098/rsta.1983.0089}
  {\path{doi:10.1098/rsta.1983.0089}}.

\bibitem{Polievktov:1969}
N.~M. Polievktov-Nikoladze, {Non-Einsteinian gravitaional equations},
  Sov.~Phys.~JETP 25 (1967) 904--909, [Zh.~Eksp.~Teor.~Fiz. {\bf 52},
  1360--1367 (1967)].

\bibitem{Ruzmaikina:1969}
T.~V. Ruzmaikina, A.~A. Ruzmaikin, {Quadratic corrections to the Lagrangian
  density of the gravitational field and the singularity}, Sov.~Phys.~JETP 30
  (1970) 372--374, [Zh.~Eksp.~Teor.~Fiz. {\bf 57}, 680--685 (1969)].

\bibitem{Ginzburg:1971}
V.~L. Ginzburg, D.~A. Kirzhnits, A.~A. Lyubushin, {The role of quantum
  fluctuations of the gravitational field in general relativity theory and
  cosmology}, Sov.~Phys.~JETP 30 (1971) 242--246, [Zh.~Eksp.~Teor.~Fiz. {\bf
  60}, 451--459 (1971)].

\bibitem{Starobinsky:1980te}
A.~A. Starobinsky, {A new type of isotropic cosmological models without
  singularity}, Phys.~Lett.~B 91 (1980) 99--102.
\newblock \href {https://doi.org/10.1016/0370-2693(80)90670-X}
  {\path{doi:10.1016/0370-2693(80)90670-X}}.

\bibitem{Weinberg:1983}
S.~Weinberg, {Overview of theoretical prospects for understanding the values of
  fundamental constants}, Phil.~Trans.~R.~Soc.~Lond.~A 310 (1983) 249--252.
\newblock \href {https://doi.org/10.1098/rsta.1983.0086}
  {\path{doi:10.1098/rsta.1983.0086}}.

\bibitem{Whitt:1984pd}
B.~Whitt, {Fourth-order gravity as general relativity plus matter}, Phys.~Lett.
  B145 (1984) 176.
\newblock \href {https://doi.org/10.1016/0370-2693(84)90332-0}
  {\path{doi:10.1016/0370-2693(84)90332-0}}.

\bibitem{Starobinsky:1987zz}
A.~A. Starobinsky, H.-J. Schmidt, {On a general vacuum solution of fourth-order
  gravity}, Class.~Quant.~Grav. 4 (1987) 695--702.
\newblock \href {https://doi.org/10.1088/0264-9381/4/3/026}
  {\path{doi:10.1088/0264-9381/4/3/026}}.

\bibitem{Maeda:1985bq}
K.-i. Maeda, {Stability and attractor in a higher-dimensional cosmology. I},
  Class. Quant. Grav. 3 (1986) 233.
\newblock \href {https://doi.org/10.1088/0264-9381/3/2/017}
  {\path{doi:10.1088/0264-9381/3/2/017}}.

\bibitem{Maeda:1987xf}
K.-i. Maeda, {Inflation as a transient attractor in $R^{2}$ cosmology}, Phys.
  Rev. D 37 (1988) 858.
\newblock \href {https://doi.org/10.1103/PhysRevD.37.858}
  {\path{doi:10.1103/PhysRevD.37.858}}.

\bibitem{Berkin:1990nu}
A.~L. Berkin, K.-i. Maeda, {Effects of R**3 and R box R terms on R**2
  inflation}, Phys. Lett. B 245 (1990) 348--354.
\newblock \href {https://doi.org/10.1016/0370-2693(90)90657-R}
  {\path{doi:10.1016/0370-2693(90)90657-R}}.

\bibitem{Capozziello:1993xn}
S.~Capozziello, F.~Occhionero, L.~Amendola, {The Phase space view of inflation:
  II. Fourth order models}, Int.~J.~Mod.~Phys. D1 (1993) 615--639.
\newblock \href {https://doi.org/10.1142/S0218271892000318}
  {\path{doi:10.1142/S0218271892000318}}.

\bibitem{Ketov:2012jt}
S.~V. Ketov, A.~A. Starobinsky, {Inflation and non-minimal scalar-curvature
  coupling in gravity and supergravity}, JCAP 08 (2012) 022.
\newblock \href {http://arxiv.org/abs/1203.0805} {\path{arXiv:1203.0805}},
  \href {https://doi.org/10.1088/1475-7516/2012/08/022}
  {\path{doi:10.1088/1475-7516/2012/08/022}}.

\bibitem{Belinskii:1985}
V.~Belinskii, L.~Grishchuk, Y.~Zel'dovich, I.~Khalatnikov, {Inflationary stages
  in cosmological models with a scalar field}, Sov.~Phys.~JETP 62 (1985)
  195--203, [Zh.~Eksp.~Teor.~Fiz.~ {\bf 89} (1985) 346-360].

\bibitem{Belinskii:1987}
V.~Belinskii, I.~Khalatnikov, {On the generality of inflationary solutions in
  cosmological models with a scalar field}, Sov.~Phys.~JETP 66 (1987) 441--449,
  [Zh.~Eksp.~Teor.~Fiz. {\bf 93} (1987) 785-799].

\bibitem{Hawking:1971bv}
S.~Hawking, {Stable and generic properties in general relativity}, Gen. Rel.
  Grav. 1 (1971) 393--400.
\newblock \href {https://doi.org/10.1007/BF00759218}
  {\path{doi:10.1007/BF00759218}}.

\bibitem{Barrow:1983rx}
J.~D. Barrow, A.~C. Ottewill, {The Stability of General Relativistic
  Cosmological Theory}, J. Phys. A 16 (1983) 2757.
\newblock \href {https://doi.org/10.1088/0305-4470/16/12/022}
  {\path{doi:10.1088/0305-4470/16/12/022}}.

\bibitem{Szydlowski:1984}
M.~Szydlowski, M.~Heller, Z.~Golda, {Structural Stability Properties of
  Friedman Cosmology}, Gen. Rel. Grav. 16 (1984) 877.
\newblock \href {https://doi.org/10.1007/BF00762940}
  {\path{doi:10.1007/BF00762940}}.

\bibitem{Coley:1992}
A.~A. Coley, R.~K. Tavakol, {Fragility in cosmology}, Gen. Rel. Grav. 24 (1992)
  835–855.
\newblock \href {https://doi.org/10.1007/BF00759090}
  {\path{doi:10.1007/BF00759090}}.

\bibitem{Kokarev:2008ba}
S.~S. Kokarev, {Structural instability of Friedmann--Robertson--Walker
  cosmological models}, Gen. Rel. Grav. 41 (2009) 1777--1794.
\newblock \href {http://arxiv.org/abs/0810.5080} {\path{arXiv:0810.5080}},
  \href {https://doi.org/10.1007/s10714-008-0748-8}
  {\path{doi:10.1007/s10714-008-0748-8}}.

\bibitem{Andronov:1937}
A.~A. Andronov, L.~S. Pontryagin, Grubyye sistemy, Dokl. Akad. Nauk SSSR 14
  (1937) 247--251, [english translation : Lev Semenovich Pontryagin, in Russian
  Mathematicians in the 20th Century, eds. Ya. G. Sinai, World Scientific
  Publishing Co. Pte. Ltd, 2003, pp. 345-366].
\newblock \href {https://doi.org/10.1142/9789812779212_0015}
  {\path{doi:10.1142/9789812779212_0015}}.

\bibitem{Thom:book}
R.~Thom, Structural Stalility and Morphogenesis: An Outline of a General Theory
  of Models, Advanced Book Classics, Westview Press, Reading, Mass., 1989.

\bibitem{Donoghue:1994dn}
J.~F. Donoghue, {General relativity as an effective field theory: The leading
  quantum corrections}, Phys.Rev. D50 (1994) 3874--3888.
\newblock \href {http://arxiv.org/abs/gr-qc/9405057}
  {\path{arXiv:gr-qc/9405057}}, \href
  {https://doi.org/10.1103/PhysRevD.50.3874}
  {\path{doi:10.1103/PhysRevD.50.3874}}.

\bibitem{Donoghue:1995cz}
J.~F. Donoghue,
  \href{http://alice.cern.ch/format/showfull?sysnb=0214301}{{Introduction to
  the effective field theory description of gravity}}, in: {Advanced School on
  Effective Theories Almunecar, Spain, June 25-July 1, 1995}, 1995.
\newblock \href {http://arxiv.org/abs/gr-qc/9512024}
  {\path{arXiv:gr-qc/9512024}}.
\newline\urlprefix\url{http://alice.cern.ch/format/showfull?sysnb=0214301}

\bibitem{ohanlon:1972}
J.~O'Hanlon, {Mach's principle and a new gauge freedom in Brans-Dicke theory},
  J. Phys. A: Gen. Phys. 5 (1972) 803.
\newblock \href {https://doi.org/10.1088/0305-4470/5/6/005}
  {\path{doi:10.1088/0305-4470/5/6/005}}.

\bibitem{Hrycyna:2010yv}
O.~Hrycyna, M.~Szydlowski, {Uniting cosmological epochs through the twister
  solution in cosmology with non-minimal coupling}, JCAP 12 (2010) 016.
\newblock \href {http://arxiv.org/abs/1008.1432} {\path{arXiv:1008.1432}},
  \href {https://doi.org/10.1088/1475-7516/2010/12/016}
  {\path{doi:10.1088/1475-7516/2010/12/016}}.

\bibitem{Hrycyna:2015eta}
O.~Hrycyna, M.~Szydlowski, {Cosmological dynamics with non-minimally coupled
  scalar field and a constant potential function}, JCAP 11 (2015) 013.
\newblock \href {http://arxiv.org/abs/1506.03429} {\path{arXiv:1506.03429}},
  \href {https://doi.org/10.1088/1475-7516/2015/11/013}
  {\path{doi:10.1088/1475-7516/2015/11/013}}.

\bibitem{Hrycyna:2020jmw}
O.~Hrycyna, {The non-minimal coupling constant and the primordial de Sitter
  state}, Eur.~Phys.~J.~C 80 (2020) 817.
\newblock \href {http://arxiv.org/abs/2008.00943} {\path{arXiv:2008.00943}},
  \href {https://doi.org/10.1140/epjc/s10052-020-8397-5}
  {\path{doi:10.1140/epjc/s10052-020-8397-5}}.

\bibitem{Kerachian:2019tar}
M.~Kerachian, G.~Acquaviva, G.~Lukes-Gerakopoulos, {Classes of nonminimally
  coupled scalar fields in spatially curved FRW spacetimes}, Phys. Rev. D
  99~(12) (2019) 123516.
\newblock \href {http://arxiv.org/abs/1905.08512} {\path{arXiv:1905.08512}},
  \href {https://doi.org/10.1103/PhysRevD.99.123516}
  {\path{doi:10.1103/PhysRevD.99.123516}}.

\bibitem{Jarv:2021qpp}
L.~J\"arv, A.~Toporensky, {Global portraits of nonminimal inflation}, Eur.
  Phys. J. C 82~(2) (2022) 179.
\newblock \href {http://arxiv.org/abs/2104.10183} {\path{arXiv:2104.10183}},
  \href {https://doi.org/10.1140/epjc/s10052-022-10124-3}
  {\path{doi:10.1140/epjc/s10052-022-10124-3}}.

\bibitem{Hrycyna:2021yad}
O.~Hrycyna, {A new generic and structurally stable cosmological model without
  singularity}, Phys.~Lett.~B 820 (2021) 136511.
\newblock \href {http://arxiv.org/abs/2105.02815} {\path{arXiv:2105.02815}},
  \href {https://doi.org/10.1016/j.physletb.2021.136511}
  {\path{doi:10.1016/j.physletb.2021.136511}}.

\bibitem{Perko:book}
L.~Perko, Differential Equations and Dynamical Systems, 3rd Edition, Vol.~7 of
  Texts in Applied Mathematics, Springer-Verlag, New York, 2001.

\bibitem{Wiggins:book}
S.~Wiggins, Introduction to Applied Nonlinear Dynamical Systems and Chaos, 2nd
  Edition, Vol.~2 of Texts in Applied Mathematics, Springer-Verlag, New York,
  2003.

\bibitem{Zeldovich:1962}
{\relax Ya}.~B. Zeldovich, {The equation of state at ultrahigh densities and
  its relativistic limitations}, Sov.~Phys.~JETP 14 (1962) 1143--1147,
  [Zh.~Eksp.~Teor.~Fiz. {\bf 41}, 1609 (1961)].

\bibitem{Zeldovich:1972zz}
{\relax Ya}.~B. Zeldovich, {A hypothesis, unifying the structure and the
  entropy of the universe}, Mon.~Not.~Roy.~Astron.~Soc. 160 (1972) 1P--3P.
\newblock \href {https://doi.org/10.1093/mnras/160.1.1P}
  {\path{doi:10.1093/mnras/160.1.1P}}.

\bibitem{Yoshida:2018ndv}
D.~Yoshida, J.~Quintin, {Maximal extensions and singularities in inflationary
  spacetimes}, Class. Quant. Grav. 35~(15) (2018) 155019.
\newblock \href {http://arxiv.org/abs/1803.07085} {\path{arXiv:1803.07085}},
  \href {https://doi.org/10.1088/1361-6382/aacf4b}
  {\path{doi:10.1088/1361-6382/aacf4b}}.

\bibitem{Nomura:2021lzz}
K.~Nomura, D.~Yoshida, {Past extendibility and initial singularity in
  Friedmann-Lema\^\i{}tre-Robertson-Walker and Bianchi I spacetimes}, JCAP 07
  (2021) 047.
\newblock \href {http://arxiv.org/abs/2105.05642} {\path{arXiv:2105.05642}},
  \href {https://doi.org/10.1088/1475-7516/2021/07/047}
  {\path{doi:10.1088/1475-7516/2021/07/047}}.

\bibitem{Carloni:2004kp}
S.~Carloni, P.~K.~S. Dunsby, S.~Capozziello, A.~Troisi, {Cosmological dynamics
  of R**n gravity}, Class. Quant. Grav. 22 (2005) 4839--4868.
\newblock \href {http://arxiv.org/abs/gr-qc/0410046}
  {\path{arXiv:gr-qc/0410046}}, \href
  {https://doi.org/10.1088/0264-9381/22/22/011}
  {\path{doi:10.1088/0264-9381/22/22/011}}.

\bibitem{Amendola:2006kh}
L.~Amendola, D.~Polarski, S.~Tsujikawa, {Are f(R) dark energy models
  cosmologically viable ?}, Phys. Rev. Lett. 98 (2007) 131302.
\newblock \href {http://arxiv.org/abs/astro-ph/0603703}
  {\path{arXiv:astro-ph/0603703}}, \href
  {https://doi.org/10.1103/PhysRevLett.98.131302}
  {\path{doi:10.1103/PhysRevLett.98.131302}}.

\bibitem{Capozziello:2006dj}
S.~Capozziello, S.~Nojiri, S.~D. Odintsov, A.~Troisi, {Cosmological viability
  of f(R)-gravity as an ideal fluid and its compatibility with a matter
  dominated phase}, Phys. Lett. B 639 (2006) 135--143.
\newblock \href {http://arxiv.org/abs/astro-ph/0604431}
  {\path{arXiv:astro-ph/0604431}}, \href
  {https://doi.org/10.1016/j.physletb.2006.06.034}
  {\path{doi:10.1016/j.physletb.2006.06.034}}.

\end{thebibliography}

\end{document}